\documentclass [longauth]{aa}
\usepackage[]{natbib}
\bibpunct{(}{)}{;}{a}{}{,}
\usepackage{subfig}
\usepackage{graphicx}
\usepackage{epsf}
\usepackage{rotating}
\usepackage{graphics}
\usepackage{txfonts}
\usepackage{amssymb}
\usepackage[breaklinks=f]{hyperref}
\newcommand{\corot}{{\it CoRoT}}
\newcommand{\kepler}{{\it Kepler}}
\newcommand{\plato}{\emph{PLATO}}
\newcommand{\tess}{\emph{TESS}}
\newcommand{\cheops}{\emph{CHEOPS}}
\newcommand{\ktwo}{{\it K2}}

\newcommand{\kms}{km\,s$^{-1}$}
\newcommand{\ms}{m\,s$^{-1}$}
\newcommand{\msd}{m\,s$^{-1}$\,d$^{-1}$}

\newcommand{\smass}[1][]{ $ 1.0 \pm 0.07 $ #1} 
\newcommand{\dsepmass}[1][]{ $ 0.88 \pm 0.02 $ #1}
\newcommand{\gsepmass}[1][]{ $ 0.86 \pm 0.04 $ #1}
\newcommand{\sradius}[1][]{ $1.4\pm 0.14 $ #1}
\newcommand{\dsepradius}[1][]{ $1.23\pm 0.1 $ #1}
\newcommand{\gsepradius}[1][]{ $1.28\pm 0.36 $ #1}
\newcommand{\stemp}[1][]{ $ 5730 \pm 50 $ #1 }
\newcommand {\Lsun}[1][]{ $1.9_{-0.4}^{+0.4} $ #1 }
\newcommand{\Tzerob}[1][]{$7067.9704 _{ - 0.0039 } ^ { + 0.0044 } $#1} 
\newcommand{\Pb}[1][]{$5.35117 \pm 0.00055$ #1} 
\newcommand{\bb}[1][]{$0.633 _{ - 0.128 } ^ { + 0.091 }$ #1}  
\newcommand{\arb}[1][]{$9.59 _{ - 0.95 } ^ { + 0.98 }$ #1}   
\newcommand{\rrb}[1][]{$0.01255 _{ - 0.00048 } ^ { + 0.00050  }$ #1}
\newcommand{\kb}[1][]{$3.1 \pm 1.4  $ #1}  
\newcommand{\ib}[1][deg]{$ 86.2 \pm 1.0   $ #1} 
\newcommand{\ab}[1][AU]{$ 0.0621 _{ - 0.0085  } ^ {+ 0.0092 }$ #1}
\newcommand{\densb}[1][]{$ 0.51 \pm 0.16$ #1}
\newcommand{\mpb}[1][$M_{\oplus}$]{$8.6 \pm 3.9$ #1} 
\newcommand{\rpb}[1][$R_{\oplus}$]{$1.9 \pm 0.2$ #1} 
\newcommand{\rhob}[1][]{$6.6_{-3.2}^{+4.5}$ #1}

\newcommand{\Tequib}[1][K]{$ 1309 _{ - 63  } ^ {+ 71 }$  #1}    
\newcommand{\ttotb}[1][]{$ 3.38 _{ - 0.10  } ^ {+ 0.11 }$  #1}
\newcommand{\tinegb}[1][]{$ 0.069 _{ - 0.014 } ^ { + 0.019  }$ #1}

\newcommand{\uone}[1][]{ $0.38 \pm 0.08 $ #1}   
\newcommand{\utwo}[1][]{ $0.28 \pm 0.08  $ #1}

\newcommand{\velHARPSN}[1][$\mathrm{km\,s^{-1}}$]{ $ -16.2120 \pm 0.0224 $}
\newcommand{\velFIES}[1][$\mathrm{km\,s^{-1}}$]{ $ -16.3372 \pm 0.0224 $}
\newcommand{\LinearTrend}[1][$\mathrm{m\,s^{-1}\,d^{-1}}$]{$ -0.217 \pm 0.077 $ }
\newcommand{\vsini}{$v$\,sin\,$i$}   
 
\newcommand{\vmic}{$V_{\rm mic}$}
\newcommand{\vmac}{$V_{\rm mac}$}
\newcommand{\dex}{$\rm dex$}
\newcommand{\teff}{$T_{\rm eff}$}

\newcommand{\logg}{log\,{\it g$_\star$}}

\newcommand{\lsun}{L$_{\odot}$}                          
\newcommand{\Msun}{$M_{\odot}$}
\newcommand{\Rsun}{R$_{\odot}$}

\newcommand{\mearth}{$M_{\oplus}$}
\newcommand{\rearth}{$R_{\oplus}$}

\newcommand{\mstar}{$M_{\star}$}
\newcommand{\rstar}{$R_{\star}$}

\newcommand{\bjdtdb}{\ensuremath{\rm {BJD_{TDB}}}}
\newcommand{\about}{$\sim$}                       
\newcommand{\av}{A$_{V}$}

\newcommand{\vmag}{$m_{V}$}

\newcommand{\asec}{$^{\prime \prime}$}

\newcommand{\corseven}{{CoRoT-7b}}		

\newcommand{\attw}{{ATLAS12}}           

\newcommand{\targeta}{EPIC\,210894022}

\newcommand{\targetb}{{EPIC\,210894022b}}

\begin{document}
\title{\targetb\ $-$\,A short period super-Earth \\ transiting a metal poor, evolved old star}
\author{Malcolm Fridlund\inst{1,2}
\and 
Eric Gaidos \inst{3}  
\and
Oscar Barrag\'an\inst{4}
\and
Carina M. Persson\inst{2}
\and
Davide Gandolfi \inst{4}   
\and
Juan Cabrera \inst{6}
\and
Teruyuki Hirano \inst{7}
\and
Masayuki Kuzuhara \inst{19,20}
\and
Sz. Csizmadia\inst{6}
\and
Grzegorz Nowak\inst{10,11} 
\and
Michael Endl\inst{14}
 \and 
Sascha Grziwa \inst{8}
\and 
Judith Korth \inst{8}
\and 
Jeremias Pfaff\inst{13}
\and
Bertram Bitsch\inst{9}
\and
Anders Johansen\inst{9}
\and
Alexander J. Mustill\inst{9}
\and
Melvyn B. Davies\inst{9}
\and 
Hans Deeg\inst{10,11}
\and
Enric Palle\inst{10,11} 
\and
William D. Cochran\inst{14}
\and
Philipp Eigm\"uller\inst{6}
 \and
Anders Erikson\inst{6}
 \and
Eike Guenther\inst{12}
\and
Artie P. Hatzes\inst{12}
 \and
Amanda Kiilerich\inst{15}
 \and
Tomoyuki Kudo\inst{21}
 \and
 Phillip MacQueen\inst{14}
 \and
Norio Narita\inst{18,19,20}
 \and
David Nespral\inst{10,11}
 \and
Martin P\"atzold\inst{8}
 \and
Jorge Prieto-Arranz\inst{10,11}
 \and
Heike Rauer\inst{6,13}
 \and
Vincent Van Eylen\inst{1}
 }  
   \institute{Leiden Observatory, University of Leiden, PO Box 9513, 2300 RA, Leiden, The Netherlands 
          \email{fridlund@strw.leidenuniv.nl}
\and
Department of Earth and Space Sciences, Chalmers University of Technology, Onsala Space Observatory, 439 92, Onsala, Sweden
\and
Department of Geology and Geophysics, University of Hawaii at Manoa, Honolulu, HI 96822, USA
\and
Dipartimento di Fisica, Universit\'a di Torino, via Pietro Giuria 1, I-10125, Torino, Italy\label{Torino}
\and
Landessternwarte K\"onigstuhl, Zentrum f\"ur Astronomie der Universit\"at Heidelberg, K\"onigstuhl 12, D-69117 Heidelberg, Germany\label{LSW}
\and 
Institute of Planetary Research, German Aerospace Center (DLR), Rutherfordstrasse
2, D-12489 Berlin, Germany
\and
Department of Earth and Planetary Sciences, Tokyo Institute of Technology, Meguro-ku, Tokyo, Japan
\and
Rheinisches Institut f\"ur Umweltforschung an der Universit\"at zu K\"oln, Aachener Strasse 209, 50931 K\"oln, Germany
\and
Lund Observatory, Department of Astronomy and Theoretical Physics, Lund University, 22100, Lund, Sweden
\and
Instituto de Astrof'\i sica de Canarias, 38205 La Laguna, Tenerife, Spain
\and
Departamento de Astrof'\i sica, Universidad de La Laguna, 38206 La Laguna, Tenerife, Spain
\and
Th\"uringer Landessternwarte Tautenburg, Sternwarte 5, 07778 Tautenburg, Germany
\and
Center for Astronomy and Astrophysics, TU Berlin, Hardenbergstr. 36, D-10623 Berlin, Germany
\and
Department of Astronomy and McDonald Observatory, University of Texas at Austin, 2515 Speedway, Stop C1400, Austin, TX 78712, USA
\and
Stellar Astrophysics Centre, Department of Physics and Astronomy, Aarhus University, Ny Munkegade 120, DK-8000 Aarhus C, Denmark
\and
Department of Physics and Kavli Institute for Astrophysics and Space Research, Massachusetts Institute of Technology, Cambridge, MA 02139, USA
\and
Princeton University, Department of Astrophysical Sciences, 4 Ivy Lane, Princeton, NJ 08544 USA
\and
Department of Astronomy, The University of Tokyo, 7-3-1 Hongo, Bunkyo-ku, Tokyo 113-0033, Japan
\and
Astrobiology Center, NINS, 2-21-1 Osawa, Mitaka, Tokyo 181-8588, Japan
\and
National Astronomical Observatory of Japan, NINS, 2-21-1 Osawa, Mitaka, Tokyo 181-8588, Japan
\and
Subaru Telescope, National Astronomical Observatory of Japan, 650 North Aohoku Place, Hilo, HI 96720, USA
}      
   \date{January 2017}
  \abstract
{ The star \targeta\ has been identified from a light curve acquired through the K2 space mission as possibly orbited by a transiting planet.} 
{Our aim is to confirm the planetary nature of the object and derive its fundamental parameters.}
{We analyse the light curve variations during the planetary transit using packages developed specifically for exoplanetary transits. Reconnaissance spectroscopy and radial velocity observations have been obtained using three separate telescope and spectrograph combinations. The spectroscopic synthesis package SME has been used to derive the stellar photospheric parameters that were used as input to various stellar evolutionary tracks in order to derive the parameters of the system. The planetary transit was also validated to occur on the assumed host star through adaptive imaging and statistical analysis.}
{The star is found to be located in the background of the Hyades cluster at a distance at least 4 times further away from Earth than the cluster itself. The spectrum and the space velocities of \targeta\ strongly suggest it to be a member of the thick disk population. The co-added high-resolution spectra show that that it is a metal poor ([Fe/H]\,=\,$-0.53\pm0.05$\,\dex) and $\alpha$-rich somewhat evolved solar-like star of spectral type G3. We find an \teff\,= $5730\pm50$ K, \logg\,=\,$4.15\pm0.1$ cgs, and derive a radius of \rstar\ =\,$1.3\pm0.1$ \Rsun and a mass of \mstar\ =\,$0.88\pm0.02$ \Msun. The currently available  radial velocity data confirms a super-Earth class planet with a mass of \mpb\ and a radius of \rpb. A second more massive object with a period longer than about 120 days is indicated by a long term radial velocity drift.}
{The radial velocity detection together with the imaging confirms with a high level of significance that the transit signature is caused by a planet orbiting the star \targeta. This planet is also confirmed in the radial velocity data. A second more massiveobject (planet , brown dwarf or star) has been detected in the radial velocity signature. With an age of $\gtrsim\,10$ Gyrs this system is one of the oldest where planets is hitherto detected. Further studies of this planetary system is important since it contains information about the planetary formation process during a very early epoch of the history of our Galaxy.}
{}
 \keywords{planetary systems -- stars: fundamental parameters -- stars: individual: \object{EPIC 210894022} -- planets and satellites: detection -- planets and satellites: fundamental parameters -- techniques: photometric -- techniques: radial velocities -- techniques: spectroscopic}
 
  \authorrunning{Fridlund et al.}
  \titlerunning{The Super-Earth planet EPIC\,210894022b}
  \maketitle
  
%

\section{Introduction}
\label{intro}
Exoplanetary transits give valuable information about the planetary size in terms of the host star. Very high precision transit photometry, preferably carried out from space, gives us access to the orbital parameters, which combined with either radial velocity (RV) data and/or transit timing variations (TTVs) enables the measurement of the planetary fundamental parameters, most notable, the planet's radius, mass, and mean density \citep{Char2000,Henry2000,Mayor1995,Marcy1996,Ford11}. Determination of the fundamental parameters of exoplanets, and their host stars, are necessary in order to study the internal structure, composition, dynamical evolution, tidal interactions, architecture of 
systems, and the atmosphere of exoplanets  
\citep{Madhusudhan2014, Winn2015, Hatzes2016}. 

The successful \corot\ and \kepler\ space missions \citep{baglinf06,borucki10}, have found large numbers of transiting exoplanets of different types and have also led to the discovery and measurements of the fundamental parameters of the first rocky exoplanets \corseven\ and Kepler-10b \citep{leger09,queloz09,Hatzes11,Batalha2011}, as well as introduced detailed modelling to the field of exoplanetary science \citep{Moutou2013}. One of the most important results of these missions is the realisation of how diverse exoplanets are. Later discoveries, primarily by the \kepler\ mission, have led to the understanding that small and dense planets ("super-Earths") are quite common \citep{Borucki2011,Mayor2011a,Mayor2011b,Torres2015,Marcy2014a,Marcy2014b}, and that they may even have formed early in our Galaxy's evolution \citep{Campante2015}. 
 
The repurposed  \ktwo\ space mission, provides long-timeline, high-precision photometry for exoplanet and astrophysics research. It is the new name given to NASA's \kepler\ mission after the failure of one of its non-redundant reaction wheels in May 2013 which caused the pointing precision of the telescope to be non-compliant with the original mission.  \ktwo\ was resumed in early 2014 by adopting a completely different observing strategy \citep{Howell2014}. The key difference of this new strategy with respect to the original one, is that the telescope can now only be pointed towards the same field in the sky for a period of maximum of $\sim$80 days, and has to be confined to regions close to the ecliptic. \ktwo\ is thus limited instead to detect planets with much shorter orbital periods than \kepler.  \ktwo\ observes stars that are on average 2-3 magnitudes brighter than those targeted by the original \kepler\ mission \citep{Howell2014}, and in fields (designated "campaigns"), re-targeting every $\sim$80 days along the ecliptic. This entails an opportunity to gain precious knowledge on the mass of small exoplanets via ground based radial velocity follow-up observations. By observing almost exclusively brighter stars than the previous missions the quality of the necessary ground based follow-up observations (e.g., spectroscopic characterisations and radial velocity measurements) has improved significantly. 

The approximately $10,000$\,--\,$15,000$ objects observed in each field are listed in the Ecliptic Plane Input Catalog (EPIC) of the \ktwo\ mission\footnote{https://archive.stsci.edu/k2.}. The capability of \ktwo\ to detect small (down to super-Earth size) transiting planets in short period orbits around such stars has recently been demonstrated \citep{Vanderburg2015}. 
 
As part of our ongoing studies of individual exoplanetary candidates from the \ktwo\ mission, and using methods \citep{Gaidos2017} we develop for the interpretation of \ktwo\, as well as the expected \tess\ \citep{Ricker2015}, \cheops\ \citep{Broeg2013}, and \plato\ missions \citep{rauer14}, we have confirmed a short-period transiting super-Earth that together with a larger body with a significantly longer period, orbits the solar-like star \targeta\footnote{The star was a target of three programs during \ktwo\ Campaign 4, GO4007, GO4033  and GO4060.}. This star was previously designated as a false positive \citep{Crossfield2016}. As is true in this case, and as it was learned during the \corot\ mission, it is quite common that automatic analysis methods give false positives for true detections, and the evolution of the pipeline software during a space mission may motivate further analyses. It should also be stressed in this context that different algorithms may give differing results. The star is a metal poor, high velocity object indicative of an old age. Planets orbiting such stars are very rare and important since they provide information about the earliest phases of planetary formation in our Galaxy. In this paper we describe our follow-up study of this object, to confirm the planetary nature of the transits, model the evolution and age of the system, as well as the formation process.

The paper is organised in the following way: in Sect. 2 we present the \ktwo\ photometry, and in Sect. 3 we present the ground based follow-up with spectral classification and validation of the planetary signal with a calculation of the false positive probability. In Sect. 4 and 5 we classify the host star kinematically, determine its distance and derive the stellar mass, the radius and age of the system. In Sect. 6 we then carry out the transit and radial velocity curve modelling and determine the exoplanetary physical parameters, the results of which make it increasingly probable that there is a second body in this system. In Sect. 7 we model the orbital dynamics of the system and finally, in Sect. 8 we discuss and summarise the results.
\begin{table}[!t]
\caption{Main identifiers, optical and infrared magnitudes, and proper motion of \targeta. \label{Tab:StarIdentifiers}}
\begin{center}
\begin{tabular}{lcc} 
\hline
\hline
\noalign{\smallskip}
Parameter & Value &  Source$^{1}$ \\
\noalign{\smallskip}
\hline
\noalign{\smallskip}
\multicolumn{3}{l}{\emph{Main Identifiers}} \\
\noalign{\smallskip}
EPIC & 210894022  & EPIC \\
UCAC2 & 39261536 & UCAC2 \\
UCAC4 & 557-008366 & UCAC4 \\
2MASS & 03593351+2117552 & 2MASS \\
\noalign{\smallskip}
\hline
\noalign{\smallskip}
\multicolumn{3}{l}{\emph{Equatorial coordinates}} \\
\noalign{\smallskip}
$\alpha$(J2000.0) & $03^\mathrm{h}\,59^{\mathrm{m}}\,33.541^{\mathrm{s}}$ & UCAC4 \\
$\delta$(J2000.0) & $21^{\circ}\,17^\prime\,55.27{\arcsec}$ & UCAC4 \\
\noalign{\smallskip}
\hline
\noalign{\smallskip}
\multicolumn{3}{l}{\emph{Magnitudes}} \\
$B$ & 11.796$\pm$0.030 & EPIC \\
$V$ & 11.137$\pm$0.040 & EPIC \\
$g$ & 11.437$\pm$0.040 & EPIC \\
$r$ & 10.876$\pm$0.020 & EPIC \\
$J$ & 9.768$\pm$0.023 & 2MASS \\
$H$ & 9.477$\pm$0.025 & 2MASS \\
$K$ & 9.377$\pm$0.021 & 2MASS \\
$W1$ & 9.321$\pm$0.023 & AllWISE \\
$W2$ & 9.347$\pm$0.021 & AllWISE \\
$W3$ & 9.213$\pm$0.034 & AllWISE \\
$W4$ & 8.847$\pm$0.509 & AllWISE \\
\noalign{\smallskip}
\hline
\noalign{\smallskip}
\multicolumn{3}{l}{\emph{Proper motions}} \\
$\mu_{\alpha} \cos \delta$ (mas \ yr$^{-1}$) & $ 122.7 \pm 2.2 $ & UCAC4 \\
$\mu_{\delta} $ (mas \ yr$^{-1}$)            & $ -35.3 \pm 1.4 $ & UCAC4 \\
\noalign{\smallskip}
\hline
\end{tabular}
\tablefoot{\tablefoottext{1}{Values of fields marked with EPIC are taken from the Ecliptic Plane Input Catalog, available at \url{http://archive. stsci.edu/k2/epic/search.php}. Values marked with UCAC2, UCAC4, 2MASS, and AllWISE are from \citet{Zacharias2004}, \citet{Zacharias2013}, \citet{Cutri2003}, and \citet{Cutri2014}, respectively.}}
\end{center}
\end{table}

\section{\ktwo\ photometry of the transit signal}
\label{transit}
Observations of the \ktwo\ Field 4 took place between February 7 and April 23, 2015. This campaign included the \object{Hyades}, \object{Pleiades}, and \object{NGC\,1647} clusters. This was by intention and most selected targets were members of these clusters. A total of 15\,847 long cadence (30 minute integration time) and 122 short cadence (1 minute integration time) targets were observed, and the data were made publicly available on September 4, 2015.
 
The part of the light curve containing the actual primary (and possibly also a secondary) transit provide significant information about both the transiting object and the host star \citep{seager03}. The actual light curve is, however, contaminated with noise caused by a number of instrumental and natural effects and needs to be processed before it can be interpreted. We used two different and independent methods to produce cleaned and interpretable light curves for all 15\,969 targets. The first technique follows the methodology outlined in \cite{Grziwa2016}. The \ktwo\ target pixel files were analyzed for stellar targets and a mask for each target was calculated and assigned. After the light curve extraction, disturbances produced by the drift\footnote{This drift is caused by the fact that the operation of the \kepler\ spacecraft using only two reaction wheels, requires using a combination of carefully balanced solar radiation pressure together with the fine adjustment thrusters in order to stabilize the spacecraft around the third axis. This results in a periodic rotation of the spacecraft about the bore sight of the telescope \citep{Howell2014}.} of the telescope over the sky were corrected by computing the rotation of the telescope's CCDs\footnote{The focal plane of \ktwo\ is equipped with an array of 21 individual CCD's covering an area of $\sim$116~deg$^2$ on the sky}. After corrections we then used the \texttt{EXOTRANS} based pipeline \citep{Grziwa2012} in order to separate stellar variability and discontinuities and to search for transit signals in the resulting light curves. 

In the second method, we used circular apertures to extract the light curves. An optimal aperture size was selected in order to minimize the noise. The background was estimated by calculating the median value of the target pixel file after the exclusion of all pixels brighter than a threshold value that may belong to a source. The resulting light curves were de-correlated using the movement of the centroid as described in \cite{VandJohn2014}. For more details we refer to \cite{Johnson2016}. We then used the  D\'etection Sp\'ecialise\'e de Transits (\texttt{DST}) algorithm \citep{Cabrera2012}, originally developed for the \corot\ mission to search for transit signals in the resulting light curves. 

Both the \texttt{EXOTRANS} and \texttt{DST} algorithms have been applied extensively to both \corot\ \citep{Carpano2009, Cabrera2009, Fridlund2010, Erikson2012, Carone2012, Cavarroc2012} and \kepler\ data \citep{Cabrera2014, Grziwa2016}. These transit detection algorithms search for a pattern in the data and use statistics to decide if a signal is present in the data or not, for example box-fitting Least Squares (\texttt{BLS}) algorithms \citep{Kovacs2002}. \texttt{DST} uses an optimized transit shape, with the same number of free parameters as \texttt{BLS}, and an optimized statistic for signal detection. \texttt{EXOTRANS} uses a combination of the wavelet based filter technique \texttt{VARLET} \citep{Grziwa2016} and \texttt{BLS}. \texttt{VARLET} was originally developed to remove or reduce the impact of stellar variability and discontinuities in the light curves of the \corot\ mission.

\begin{figure}[!t]
 \centering
   \includegraphics[width=\linewidth]{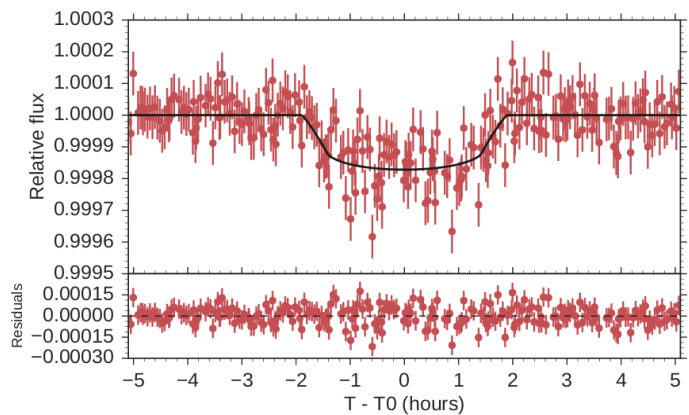}
   \caption{Transit light curve folded to the orbital period of \targetb\ and residuals. The red points mark the \ktwo\ photometric data and their error bars. The integration time of the\ktwo\ data is 30 minutes. The solid line marks the best-fitting transit model super-sampled using ten sub-samples per \ktwo\ exposure to reduce the effects from the long integration.}
      \label{F1}
 \end{figure}

When applied, both \texttt{EXOTRANS} and \texttt{DST} resulted in the discovery of a shallow transit signature in the light curve of the star designated  \targeta\ occurring every $\sim$5.35~days. The depth of the signal (\about~0.014\%), shown in Fig.~\ref{F1}, is compatible with a super-Earth-size planet transiting a solar-like star. Table~\ref{Tab:StarIdentifiers} lists the main designations, optical and infrared magnitudes, and proper motion of \targeta. The detection and characterisation of the planet were then confirmed using \cite{VandJohn2014}\footnote{https://www.cfa.harvard.edu/~avanderb/k2.html} and EVEREST lightcurves \citep{Luger2016}. Together with \texttt{EXOTRANS} and \texttt{DST}, we  obtained consistent parameters (e.g., period, depth, duration) within the uncertainties.

The analysis of the light curve extracted with Vanderburg's pipeline revealed a transit-like feature close to phase 0.5 in the folded light curve with a  significance of 3.6 sigma. Depending on the circumstances, the presence of secondary eclipses in the folded light curve of a planetary candidate can be a clear sign of contamination by background eclipsing binaries. Ruling out the presence of such secondary eclipses is a mandatory step in the photometric confirmation of planetary candidates. It was found that the transit-like feature was not consistent with the expected duration and dilution factor of a secondary eclipse by a background eclipsing binary. The duration and depth of the transit-like feature actually depended on the binning chosen in the folding process, which is typically not the case for genuine astrophysical signals. We concluded that the transit-like feature was either some residual of correlated noise in the light curve or simply a statistical fluctuation without astrophysical origin.

\section{Ground-based follow-up}
\subsection{High resolution spectroscopy}
\label{Sect:RV}

In November 2015 we obtained 4 reconnaissance high-resolution ($R \approx 60\,000$) spectra of \targeta\ using the {\bf Coud\'e} Tull spectrograph \citep{Tull1995} at the 2.7-m telescope at McDonald Observatory (Texas, USA). The spectra have a signal-to-noise ratio (SNR) of $\sim$25-40 per resolution element at 5500\,\AA. We reduced the data using standard \texttt{IRAF} routines and derived preliminary spectroscopic parameters using the code \texttt{Kea} \citep{Endl2016} and radial velocities via cross-correlation with the RV standard star \object{HD\,50692}. The results from all 4 spectra are nearly identical and reveal 
a star with effective temperature, \teff\ = 5\,778 $\pm60$\,K , surface gravity, \logg\ = 4.19 $\pm0.2$\,dex, metallicity, [M/H] = $-0.3\pm0.1$ dex and 
a slow projected rotational velocity of $3.7\pm0.3$\,\kms. The spectra show no significant radial velocity variation at a level of $\sim$150 m/s.

\begin{table}[!t]
 \centering
 \caption{FIES and HARPS-N RV measurements of \targeta.}
 \label{Table:RV}
\begin{tabular}{lcccr}
\hline
\hline
\noalign{\smallskip}
BJD$^{1}$&   RV     & eRV & FWHM$^{1}$ & BIS$^{1}$  \\
(-2450000.0) & (\kms) & (\kms) & (\kms) & (\kms) \\
\noalign{\smallskip}
\hline
\noalign{\smallskip}
\multicolumn{4}{l}{FIES} \\
\noalign{\smallskip}
7342.501727 & -16.3994  & 0.0054 & 11.7051 & -0.0087 \\
7344.554911 & -16.3959  & 0.0062 & 11.6857 & 0.0185 \\
7345.481050 & -16.3918  & 0.0066 & 11.7134 & -0.0090 \\
7345.602200 & -16.3943  & 0.0068 & 11.7379 & 0.0114 \\
7346.471723 & -16.4020  & 0.0089 & 11.6864 & -0.0001 \\
7347.466106 & -16.4022  & 0.0056 & 11.7246 & 0.0109 \\
\noalign{\smallskip}
\multicolumn{4}{l}{HARPS-N} \\
\noalign{\smallskip}
7345.565450 & -16.2688  & 0.0037  & 6.6644 &    0.0048 \\
7345.591665 & -16.2664  & 0.0043  & 6.6678 &    0.0019 \\
7345.609883 & -16.2675  & 0.0045  & 6.6765 &    0.0037  \\
7346.583757 & -16.2714  & 0.0087  & 6.6447 & $-$0.0188  \\
7347.567664 & -16.2748  & 0.0042  & 6.6440 &    0.0084 \\
7347.588231 & -16.2758  & 0.0043  & 6.6677 & $-$0.0204 \\
7348.560423 & -16.2767  & 0.0022  & 6.6689 &    0.0053 \\
7370.540670 & -16.2758  & 0.0025  & 6.6622 &    0.0014 \\
7370.561584 & -16.2745  & 0.0026  & 6.6725 &    0.0067 \\
7371.457513 & -16.2743  & 0.0026  & 6.6627 & $-$0.0060 \\
7371.478774 & -16.2781  & 0.0020  & 6.6663 & $-$0.0006 \\
7399.323063 & -16.2791  & 0.0053  & 6.6820 & $-$0.0016 \\
\noalign{\smallskip}
\hline
\end{tabular}
\tablefoot{\tablefoottext {1}{FWHM is the full-width at half maximum and BIS is the bisector span of the cross-correlation function (CCF). Time stamps are given in barycentric Julian day in barycentric dynamical time (BJD$_\mathrm{TDB}$)}}
\end{table}

We started the high-precision RV follow-up of \targeta\ using the Fibre-fed Echelle Spectrograph \citep[FIES;][]{Frandsen1999,Telting2014} mounted at the 2.56-m Nordic Optical Telescope (NOT) of the Roque de los Muchachos Observatory (La Palma, Spain). We collected 6 high resolution spectra (R\,$\approx$\,67\,000) in November 2015, as part of the CAT observing program 35-MULTIPLE-2/15B. The exposure time was set to 2400s\,--\,3600s, leading to a SNR of 40\,--\,60 per pixel at 5500~\AA. In order to remove cosmic ray hits, we split each exposure in 3 consecutive sub-exposures of 800s\,--\,1200s. Following the observing strategy outlined in \citet{Buchhave2010} and \citet{Gandolfi15}, we traced the RV drift of the instrument by acquiring long exposure ($\mathrm{T_{exp}\approx35s}$) ThAr spectra immediately before and after the three sub-exposures. The data were reduced following \texttt{IRAF} and \texttt{IDL} routines. Radial velocities were extracted  via SNR-weighted, multi-order, cross-correlation with the RV standard star \object{HD\,50692} which was observed with the same instrument set-up as the target. 

Twelve additional high-resolution spectra (R\,$\approx$\,115\,000) were obtained with the HARPS-N spectrograph \citep{Cosentino2012} mounted at the 3.58-m Telescopio Nazionale Galileo (TNG) of Roque de los Muchachos Observatory (La Palma, Spain). The observations were performed between November 2015 and January 2016 as part of CAT and OPTICON programs 35-MULTIPLE-2/15B, 15B/79 and 15B/064. 

We set the exposure to 1800s and monitored the sky background using the second fibre. The data reduction was performed with the dedicated HARPS-N pipeline. The extracted spectra have a SNR of 20 - 60 per pixel at 5500~\AA. Radial velocities were extracted by cross-correlation with a G2 numerical mask \citep{Baranne96,Pepe02}.

The FIES and HARPS-N RVs are listed in Table~\ref{Table:RV}, along with the full-width at half maximum (FWHM) and the bisector span (BIS) of the cross-correlation function (CCF). Time stamps are given in barycentric Julian day in barycentric dynamical time (BJD$_\mathrm{TDB}$).

The FIES and HARPS-N RVs show a $\sim$2-$\sigma$ significant RV variation in phase with the \ktwo\ ephemeris, and, superimposed on a long negative linear trend ($\dot{\gamma}=-0.217\pm0.077~{\rm\,m\,s^{-1}\,d^{-1}}$  with a $\sim$3-$\sigma$ significance level.), as discussed in Sect.~\ref{Sect:Joint_Modeling}. In order to assess if the observed RV variation is caused by a distortion of the spectral line profile -- unveiling the presence of activity-induced RV variations and/or of a blended eclipsing binary system -- we searched for possible correlations between the RV and the BIS and FWHM measurements. The linear correlation coefficient between the RV and FWHM measurements is 0.14 (p-value = 0.79) for the FIES data, and $-$0.13 (p-value = 0.70) for the HARPS-N data; the correlation coefficient between 
the RV and BIS measurements is $-$0.14 (p-value = 0.79) for FIES, and 0.15 (p-value = 0.64) for HARPS-N. The lack of significant correlations suggest that the observed RV variations are Doppler shifts induced by the orbiting companions. We can therefore confirm the transiting planetary candidate with a mass of \mpb, and find support for the presence of a secondary body with a significantly longer period.

\subsection{Spectral classification}
\label{spectrum}
\begin{figure}
\centering
  \includegraphics[width=\linewidth]{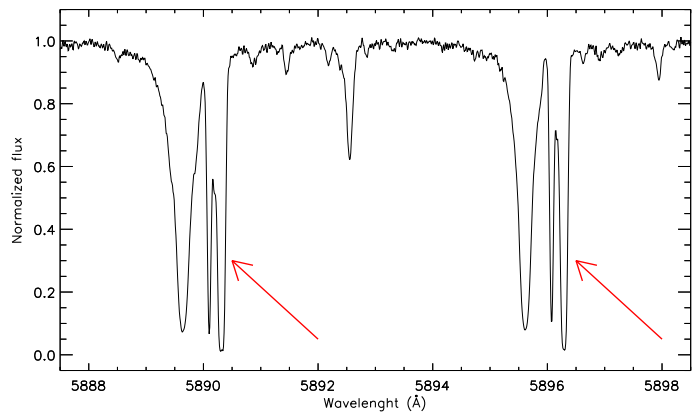}
  \caption{Spectral region encompassing the Na D doublet. The interstellar Na lines are indicated with red arrows.}
  \label{F2}
\end{figure}
 The most useful method to determine the fundamental stellar parameters (e.g. \mstar, \rstar, and the stellar age), required for the interpretation of the exoplanet data, is so far to analyse the high resolution spectra obtained in order to prepare the RV curve used for the planetary mass determination. After correcting for the RV variation, the spectra of the FIES and HARPS-N spectra were co-added to produce a high signal-to-noise ratio (SNR).  This resulted in one spectrum with SNR $\sim$120 per pixel at $5\,500$ \AA\ for the co-added FIES spectrum and another with SNR $\sim$150 at $5\,500$~\AA\ for the HARPS-N spectrum respectively. 

To determine the \teff\, the profile of either of the strong Balmer line wings is then  fitted to the appropriate stellar spectrum models \citep{fuhrmann93,axer94,fuhrmann94,fuhr97a,fuhr97b}. This fitting procedure has to be carried out carefully since the determination of the level of the adjacent continuum can be difficult for modern high-resolution Echelle spectra where each order can only contain a limited wavelength band \citep{fuhr97a}. A suitable part of the Balmer line core is excluded since this part of the line profile originate in layers above the actual photosphere and thus would be contributing to a different value of the \teff. 
\begin{table*}
\centering
\caption{\targeta\ system parameters.}
\begin{tabular}{lcrr}
\hline
\hline
\noalign{\smallskip}
Parameter & Units & Value & Comment\\
\noalign{\smallskip}
\hline
\noalign{\smallskip}
\multicolumn{3}{l}{\emph{Stellar Parameters}}\\
\noalign{\smallskip}
                         ~~~$M_{*}$\,(Spectra) &Mass (\Msun)\dotfill & \smass& \citep{Torres2010}\\
\noalign{\smallskip}
                         ~~~$R_{*}$\, (Spectra)\dotfill &Radius (\Rsun)\dotfill & \sradius&\citep{Torres2010}\\
\noalign{\smallskip}
                         ~~~$M_{*}$\, (Model)&Mass (\Msun)\dotfill & \dsepmass& DSEP Sect. \ref{Age}\\
\noalign{\smallskip}
                         ~~~$R_{*}$\,  (Model)&Radius (\Rsun)\dotfill & \dsepradius& DSEP Sect. \ref{Age}\\
\noalign{\smallskip}
                         ~~~$M_{*}$\, (Model)&Mass (\Msun)\dotfill & \gsepmass& PARAM 1.3 model Table \ref{Table:Girardi}\\
\noalign{\smallskip}
                         ~~~$R_{*}$\,  (Model)&Radius (\Rsun)\dotfill & \gsepradius& PARAM 1.3 model Table \ref{Table:Girardi}\\
\noalign{\smallskip}
                     ~~~$L_{*}$\,(Spectra)\dotfill &Luminosity (\lsun)\dotfill & \Lsun\\
\noalign{\smallskip}
                    ~~~$\rho_*$\dotfill &Density (g/cm$^3$)\dotfill & \densb  \\
\noalign{\smallskip}
                ~~~\teff\dotfill &Effective temperature (K)\dotfill & \stemp\\
\noalign{\smallskip}
~~~$\log(g_*)$\dotfill &Surface gravity (cgs) - Spectroscopy only      \dotfill & $4.15\pm0.1$\\                
\noalign{\smallskip}
                      ~~~$[$Fe/H$]$\dotfill &Iron abundance\dotfill & $-0.53\pm0.05$\\
\noalign{\smallskip}
   ~~~Age\dotfill &Gyrs\dotfill & $10.770 \pm 1.450$ & PARAM 1.3 model Table \ref{Table:Girardi}\\
\noalign{\smallskip} 
~~~Distance\dotfill & pc\dotfill&210 $\pm$20&PARAM 1.3 model Table \ref{Table:Girardi}\\
\noalign{\smallskip}                  
\multicolumn{3}{l}{\emph{Transit and Orbit Parameters}}\\
\noalign{\smallskip}                
                ~~~$P$\dotfill &Period (days)\dotfill & \Pb\\
\noalign{\smallskip}                
                ~~~$T_C$\dotfill &Time of transit (\bjdtdb)\dotfill & \Tzerob\\
\noalign{\smallskip}                          
                ~~~$T_{14}$\dotfill &Total duration (hours)\dotfill & \ttotb\\
\noalign{\smallskip}                                                                                 
                ~~~$\tau$\dotfill &Ingress/egress duration (hours)\dotfill & \tinegb\\
\noalign{\smallskip}                                
                ~~~$b$\dotfill &Impact Parameter\dotfill & \bb\\
\noalign{\smallskip}
                ~~~$i$\dotfill &Inclination (degrees)\dotfill & \ib \\
\noalign{\smallskip}
                ~~~$e$\dotfill &Eccentricity \dotfill & 0 (fixed) \\
\noalign{\smallskip}                                      
                ~~~$R_{P}/R_{*}$\dotfill &Radius of planet in stellar radii\dotfill & \rrb\\
\noalign{\smallskip}                
                ~~~$a/R_{*}$\dotfill &Semi-major axis in stellar radii\dotfill & \arb\\
\noalign{\smallskip}                
                ~~~$u_1$\dotfill &Linear limb-darkening coeff\dotfill & \uone\\
\noalign{\smallskip}                              
                ~~~$u_2$\dotfill &Quadratic limb-darkening coeff\dotfill & \utwo\\
\noalign{\smallskip}
\multicolumn{3}{l}{\emph{RV Parameters}}\\
\noalign{\smallskip}
                ~~~$K$\dotfill &RV semi-amplitude variation (\ms)\dotfill & \kb\\
\noalign{\smallskip}
                ~~~$\gamma_\mathrm{FIES}$\dotfill &Systemic velocity (FIES) (\kms)\dotfill & \velFIES\\
\noalign{\smallskip}
                ~~~$\gamma_\mathrm{HARPS-N}$\dotfill &Systemic velocity (HARPS-N) (\kms)\dotfill & \velHARPSN\\
\noalign{\smallskip}
                ~~~$\dot{\gamma}$\dotfill & Radial acceleration (\msd)\dotfill & \LinearTrend\\
\noalign{\smallskip}
\multicolumn{3}{l}{\emph{Planetary Parameters}}\\
\noalign{\smallskip}                
                   ~~~$R_{P}$\dotfill & Planet Radius (R$_\oplus$)\dotfill & \rpb \\
\noalign{\smallskip}                                             
                  ~~~$M_{P}$\dotfill & Planet Mass (M$_\oplus$)\dotfill   & \mpb \\
\noalign{\smallskip}                                              
                  ~~~$\rho_\mathrm{p}$\dotfill & Planet Density (g\,cm$^{-3}$) \dotfill & \rhob  \\
\noalign{\smallskip}                                              
                  ~~~$a$\dotfill &Semi-major axis (AU)\dotfill & \ab\\
\noalign{\smallskip}                
           ~~~$T_{eq}$\dotfill &Equilibrium Temperature$^{(1)}$ (K)\dotfill & \Tequib\\
\noalign{\smallskip}                
\hline
\end{tabular}
\tablefoot{\tablefoottext{1}{$T_{eq}$ is calculated assuming isotropic reradiation and a Bond albedo of zero.}}
\label{Table_Param}
\end{table*}

The analysis was then carried out as follows. We fitted the observed spectra to a grid of theoretical \attw\ model atmospheres from \cite{Kurucz2013}. We selected parts of the observed spectrum that contained spectral features that are sensitive to the required parameters. We used the empirical calibration equations for Sun-like stars from \cite{Bruntt2010b} and \cite{Doyle2014} in order to determine the micro-turbulent (\vmic) and macro-turbulent (\vmac) velocities, respectively. The projected stellar rotational velocity \vsini\ was measured by fitting the profile of about 100 clean and unblended metal lines. In order to calculate the best model that fitted the different parameters, we made use of the spectral analysis package SME \citep{vp96,vf05}. SME calculates, for a set of given stellar parameters, synthetic spectra and fits them to observed high-resolution spectra using a $\chi^2$  minimization procedure. We used SME version 4.43 and a grid of the \attw\ model atmospheres \citep{Kurucz2013} which is a set of 1-D models applicable to solar-like stars.

The final adopted values are listed in Table~\ref{Table_Param}. We report the individual abundances of some elements 
in Table~\ref{Table:Metals}. We find \teff\,= $5730\pm50$ K, \logg\,=\,$4.15\pm0.1$ cgs, and an iron abundance of [Fe/H]\,=\,$-0.53\pm0.05$~dex. \cite{Crossfield2016}~obtained a spectrum using the HIRES spectrograph and Specmatch. They find  \teff\ = $5788 \pm71\,$K and \logg\ = $4.224 \pm \,0.078$, in agreement with our values. Based on an average of the Ca, Si and Ti abundances (excluding the abundance of Mg, since that is based on just two lines), we find the [$\alpha$/Fe] = $+0.2 \pm$ 0.05~and \targeta\ is thus iron-poor and moderately $\alpha$-rich. 
  
Using the \cite{Straizys1981} calibration scale for dwarf stars, the effective temperature and \logg\ of \targeta\ defines the spectral type of this object as an early G-type. The low value of the \logg\ parameter suggest that  the star is evolving off the main sequence, indicating a high age and consistent with the high space velocities, as well as the low iron abundance.
\subsection{Validation of the transiting planet}

\subsubsection{High resolution imaging}
\label{validation1}

Transits such as \targetb, that appear to be planetary in origin, may actually be false positives arising from the diluted signal of a fainter, unresolved eclipsing binary (EB) that is either an unrelated background system or a companion to the primary star. In order to identify this potential false alarm source, we searched for faint stars close to the target in images acquired with high spatial resolution. \targeta\ was first observed on November 18, 2015 with the FastCam lucky imaging camera \citep{Oscoz2008} at the 1.52-m Carlos S\'anchez Telescope at Teide observatory, Tenerife. We acquired ten ``cubes'' of $1,000$ images through an I-band filter, each with 50 ms exposure time. Due to the 1.5\arcsec seeing and the relative faintness of the target, only four of these cubes could be processed successfully with the `shift and add' technique. Two processing attempts were made, using in one case the 1\% and in the other the 10\% of the images that have the smallest point spread function. In neither of the processed combined images, which cover an area of $\approx$\,$5\arcsec \times 5 \arcsec$ centred on \targeta, could any further stars be discerned, up to 4 magnitudes fainter than the target. 

In order to further check if an unresolved eclipsing binary mimics planetary transits, we also performed an adaptive-optics (AO) imaging with the HiCIAO instrument on the Subaru 8.2-m telescope \citep{Tamura2006,Suzuki2010} on December 31, 2015. Employing the AO188 \citep{2010SPIE.7736E..0NH} and Direct Imaging (DI) mode, we observed \targeta\ in the $H$ band with 3-point dithering. To search for possible faint companions, we set each exposure time to 15s $\times$ 10 coadds and let the target be saturated with the saturation radius being $\sim$0.08$^{\prime\prime}$. For the flux calibration, we also obtained an unsaturated image of \targeta\ with an exposure time of 1.5s $\times$ 5 coadds for each of the three dithering points using a 9.74\% neutral density (ND) filter. The total integration times were 900s for the saturated image and 22.5s for the unsaturated one. 

\begin{figure}[t]
\begin{center}
\includegraphics[width=8.5cm]{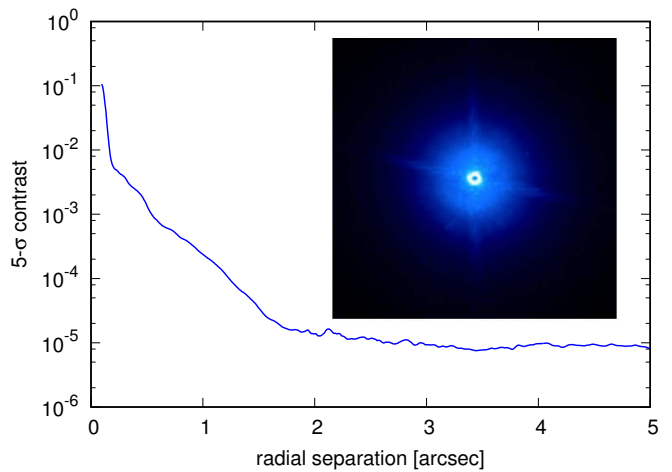}
\caption{$5\sigma$ flux contrast curve as a function of separation from \targeta. 
The inset displays the combined saturated H-band image of the target acquired with HiCIAO. The 
field-of-view is $4^{\prime\prime}\times 4^{\prime\prime}$. North is up and
East is to the left.}
\label{F3}
\end{center}
\end{figure}

We reduced the HiCIAO images following the procedure described in \cite{2013ApJ...764..183B} and \cite{2016arXiv160200638H}. The raw images  were first processed to remove the correlated read-out noises (so-called ``stripes"). The hot pixels were masked and the resulting images were flat-fielded and distortion-corrected by comparing the images of the globular cluster M5 with data taken by the Hubble Space Telescope.
 All images in each category (saturated and unsaturated) were finally aligned and median combined. The combined unsaturated image shows that the full width at half maximum (FWHM) of \targeta\ after the AO correction is $0.052^{\prime\prime}$. The images were finally aligned and median combined. With a visual inspection of the combined saturated image
(see the inset of Fig. \ref{F3}), we did not 
find any bright companion candidate up to $5^{\prime\prime}$ from the target. Two neighboring  faint objects were found to the North-East of 
\targeta\ at a separation of $\sim$8.5\asec. These objects are, however, 
only partially in the photometric aperture, and too faint (flux contrasts 
less than $4\times 10^{-5}$ in the $H$ band) to be a source of transit-like 
signals in the \ktwo\ light curve.
 

To draw a flux contrast curve around \targeta, we convolve the combined saturated image with an aperture equivalent to the FWHM of the object. The standard deviation of flux counts of the convolved image was computed within an arbitrary annulus as a function of separation from \targeta. After carrying out aperture photometry of the combined unsaturated image using an aperture radius of the FWHM of the point spread function and applying a correction for the integration times and the transmittance of the neutral density (ND) filter, we measure the $5\sigma$ contrast from \targeta. The solid line of Fig. \ref{F3} plots the measured $5\sigma$ contrast as a function of separation from the target in arcseconds and the 5-$\sigma$ contrast is $<$~than $3\times 10^{-4}$ at 1\asec. Given the transit depth of $\Delta F/ F$ = $1.8\times 10^{-4}$, we can exclude the presence of false alarm sources further than \about 1\asec away from \targeta.

\subsubsection{False Positive Probability}
\label{validation2}

To further exclude the possibility of a false positive due to a faint, blended eclipsing binary, we performed a Bayesian calculation based on the stellar background. This simulation does not include the probability that such a star is actually a binary on an eclipsing orbit, only the probability that an appropriate star is at the location of EPIC 210894022, and thus it is an upper limit on the False Positive Probability (FPP).  The procedure is described in detail in \citet{Gaidos2016} and summarized here. The Bayesian prior is based on a model of the background stellar population and the likelihoods are based on observational constraints. A background stellar population equivalent to 10 square degrees (to improve counting statistics) was constructed at the location of \targeta\ using \texttt{TRILEGAL} Version~1.6 \citep{Vanhollebeke2009}. The background was computed to $K_p = 22$, fainter than the faintest EB ($K_p \approx 20$) that could produce the signal. The likelihood for a hypothetical background star is the product of the probabilities that (a) it can produce the observed transit depth; (b) its mean density is consistent with the observed transit duration; and (c) it does not appear in our Subaru HiCIAO $H$-band imaging of the \targeta\ (Sect.~\ref{validation1}). More advanced FPP calculations can take into account the precise shape of the transit but we show that such refinement is not needed in this case.

The calculation was performed by random sampling of the synthetic background population, placing the stars in a uniformly random distribution over a region with a 15\asec\ radius centred on \targeta. Stars that exceeded the AO contrast ratio constraint (condition c) were excluded. Given the known orbital period and mean density of the synthetic star, the probability that a binary would have an orbit capable of producing the observed transit duration (condition b) was calculated assuming a Rayleigh distribution of orbital eccentricities with mean of 0.1. (Binaries on short-period orbits should quickly circularize.)\footnote{The eclipse duration calculation uses the formula for a ``small'' occulting object and so is only approximate.} To determine whether a background star could produce the observed transit signal with an eclipse depth <50\% (condition a), we determined the relative contribution to the flux of \targeta\ assuming a $7 \times 7$ pixel photometric aperture and using bilinear interpolations of the pixel response function for detector channel 48 with the tables provided in the Supplement to the \kepler\ Instrument Handbook (E. Van Cleve \& D. A. Caldwell, KSCI-19033). The calculations were performed in a series of 1000 Monte Carlo iterations and a running average used to monitor convergence.  We found a FPP of $\approx 2 \times 10^{-7}$.

We estimated the probability that the transit signal could be due to a companion EB or transiting planet system by using the 99.9\% upper limit of the stellar density derived from the fitting of the transit light curve but {\it without} spectroscopic priors. This calculates a {\it minimum} mass and radius, and by using a stellar isochrone, the absolute brightness of a hypothetical companion with the same age and metallicity as \targeta. The contrast ratio between the hypothetical stellar companion and \targeta\ can then be established via the photometric distance. 

We then used an 11.5 Gyr, [Fe/H]\,=\,-0.5 isochrone (see Sect. \ref{Age}) generated by the Dartmouth Stellar Evolution program  \citep{Dotter2008} to put lower limits on the companion effective temperature and mass, (\teff\ $>$\,5\,900\,K and \mstar\ $>$ 0.79\,\Msun) and faint limits on the magnitudes, $K_\mathrm{s}$\,$<$\,10.3 and a \kepler\ $K_\mathrm{P}$\,<\,$11.5$  using a photometric distance of 230~pc. The predicted $K$-band contrast is $<$\,0.9 magnitudes and the AO imaging we performed by Subaru-ICRS (Sect. \ref{validation1}) limits any such companion to within 0.095\asec (Fig. \ref{F3}) or about 22~AU. Such a companion would have a typical projected RV difference of at least a few \kms\ and because of the relatively modest contrast, we would have expected to resolve a second set of lines in our FIES and HARPS-N spectra, which we do not. If the companion exists and hosts the transiting object, the object must be {\it smaller} than our estimate (and thus still a planet), because the star is hotter and thus its surface brightness is higher than \targeta.
\section{The star, its distance and space velocities}
\label{star}
The object \targeta\ is a relatively bright (Table~\ref{Tab:StarIdentifiers}) star. Based on colours and proper motion measurements, \cite{Pels1975} suggested that \targeta\ is a G0 star and probably a member of the \object{Hyades} open cluster. \cite{Griffin1988} found, based on the proper motions the object to be a likely member of the Hyades, but with incompatible photometry and radial velocities. The final conclusion of those authors was that the star is not a member of the cluster. Our observations and analysis is definitely not compatible with Hyades membership. Instead we find an old, low metallicity, early G-type star (Sect.~\ref{spectrum}). The low iron abundance of $-0.53\pm0.05$ dex is not in agreement with measurements of the \object{Hyades} stars, and the apparent magnitude, \vmag\ is also not consistent with that expected for a main sequence early G star in the Hyades cluster. Radial velocity measurements of \targeta\ ($-16.3$~\kms) also support that it is not a \object{Hyades} star, since such stars on average have radial velocities of about +40 \kms. 

Considering the \vmag\ = 11.137 mag and colour index $B-V=0.659$~mag, and assuming no or very little reddening and a main sequence star of (bolometric) absolute 
magnitude $M_\mathrm{V} = 4.75$~mag, indicative of an early G-type main sequence star, we find a lower limit to the distance of $\sim$190~pc.

Figure~\ref{F2} shows our HARPS spectrum of the Na D doublet of \targeta\ where three separate components are clearly seen in each Na line: the stellar absorption profile and two (overlapping) interstellar absorption lines at different radial velocities. This is also a strong indication that the star must have a distance much larger than the \object{Hyades} cluster (45 pc).  We can correct the observed $B-V = 0.659 \pm 0.05$~for reddening using the absorption by the intervening neutral Na I along the line of sight as a measure, and the relationship between the total equivalent width of Na I absorption in both the D1 and D2 resonant lines ($0.50\pm0.05 \AA$) and E(B-V) reddening by \citet{Poznanski2012}. This relation predicts $E(B-V) = 0.055 \pm 0.014$, corresponding to an \av\ of $0.17\pm0.04$ slightly less than the upper limit one would expect from the H I column density map of \citet{Schlafly2011} of 0.18.  
  
We can also estimate the interstellar reddening towards \targeta\ following the method outlined in \cite{gandolfi08}. Briefly, we assume $R_V = 3.1$ and adopt an extinction law \citep{Cardelli1989}. We fit the spectral energy distribution using synthetic colours calculated "ad hoc" from the BT-NEXTGEN low resolution model spectrum \citep{Allard2011} with the same parameters as we find for the star (see Sect.~\ref{spectrum}), resulting in a value for \av\ of $0.15\,\pm$0.03 magnitude, similar to what we find from Na D lines. 
  
An \av\ of $0.15$ would be consistent with a distance of 210 pc if the star has the same absolute (bolometric) magnitude as the Sun. It appears, however, from our spectroscopic analysis that the star is somewhat evolved (\logg\ \about 4.15) and therefore brighter. Using the stellar parameters derived from our high resolution high signal-to-noise spectroscopy (see Sect.~\ref{spectrum}) we have \teff\ = \stemp K, which is representative of  spectral type of G3. If we then apply the equations for \mstar\ and \rstar\ derived empirically by \cite{Torres2010} we can derive an upper limit to the intrinsic luminosity of 1.9 \lsun. Using the reddening derived above this translates into a maximum distance of \about~230 pc. We therefore conclude that the distance to this object is 190 pc to 230 pc with a most likely distance of 210 $\pm$20 pc. Applying that distance to the velocity components of the star, see table \ref{Tab:StarIdentifiers}, demonstrates that \targeta\ is a very fast moving object, quite similar to the object \object{Kepler-444} studied by \cite{Campante2015}. Assuming a distance of 210 pc, we find the individual velocities with respect to the local standard of rest are ($U_{LSR}, V_{LSR}, W_{LSR}$) = ($130.6\pm2.6$,$-35.2\pm1.5$,$-16.3\pm0.5$ \kms). Correcting for the Sun's peculiar motion this is equivalent to a space velocity of 143.8 $\pm$3 \kms~almost the same as the peculiar velocity found for Kepler-444. Contrary to that object \targeta\ being of higher mass, is evolving, and therefore presumably an old object. Based on the kinematics of \targeta\, and following \cite{Reddy2006} and \cite{Sperauskas2016} we can calculate the probabilities of membership in the different populations of the Galaxy. We find that these are: 
\begin{itemize}
\item Thick disk = 96.2\%
\item Halo = 3.8\%
\item Thin disk $<$ 0.1\%
\end{itemize}

Kinematically, therefore, it is most likely that \targeta\ belongs to the thick disk population.

\begin{table}[!t]
 \centering
 \caption{Individual abundances derived assuming the effective temperature and surface gravity listed in Table~\ref{Table_Param}. All values are relative to the solar abundance.}
 \label{Table:Metals}
\begin{tabular}{lr}
\hline
\hline
Parameter  &   Value (dex)    \\
 \\
\hline
$[$Fe/H$]$ & $-0.53 \pm 0.05$ \\
$[$Ni/H$]$ & $-0.5 \pm 0.1$ \\
$[$Ca/H$]$ & $-0.2 \pm 0.1$ \\
$[$Na/H$]$ & $-0.3 \pm 0.1$ \\
$[$Ti/H$]$ & $-0.3 \pm 0.1$ \\
$[$Si/H$]$ & $-0.3 \pm 0.1$ \\
$[$Mg/H$]$ & $-0.05 \pm 0.1$\\
\hline
\end{tabular}
\end{table}
\section{The stellar mass, radius and age of the system}
\label{Age}

%

\begin{figure}[t]
 \centering
   \includegraphics[width=\linewidth]{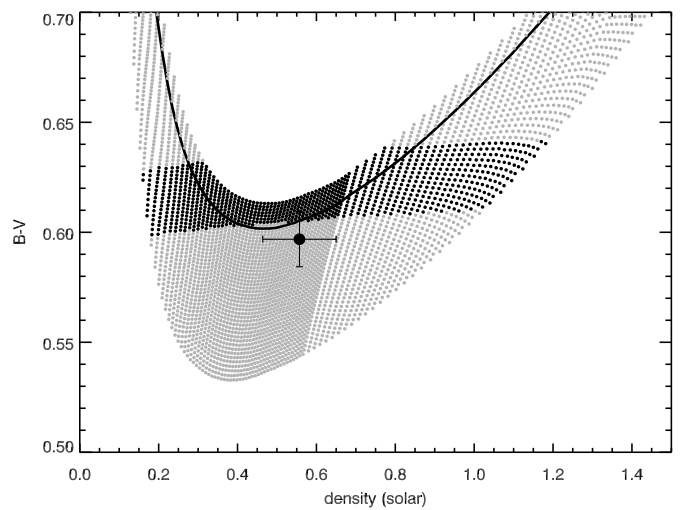}
   \caption{$B-V$ vs. stellar density (solar units).  The point is EPIC 210894022 with $B-V$ corrected for reddening based on the interstellar Na I absorption in the star's spectrum.  Dartmouth Stellar Evolution Program (DSEP)  \citep{Dotter2008} isochrones for 8-13~Gyr-old stars with [Fe/H]\,=\,-0.5 and [$\alpha$/Fe]\,=\,+0.4 are plotted; heavy points are 
   for those with \teff\ within 50 K of the spectroscopic value of 5\,730 K.  The solid line is the 12.5 Gyr old isochrone which minimizes the $\chi^2$ function.}
   \label{F7}
 \end{figure}

 We can infer stellar parameters, including age, by comparing the observed parameters to those predicted by the Dartmouth Stellar Evolution Program (DSEP)  \citep{Dotter2008}.  We selected isochrones for [Fe/H] = -0.5 and [$\alpha$/Fe] = +0.2 and +0.4, and compared predicted parameters vs. observed $B-V$, density $\rho_*$, and spectroscopic \teff\ and \logg, via a standard $\chi^2$ function, which is minimized. Applying the correction for reddening quoted in Sect.~\ref{star}, we plot the reddening-corrected $B-V$  versus the density in Fig. \ref{F7} and compared to the DSEP predictions for [Fe/H] = -0.5 and [$\alpha$/Fe]=+0.4.  
The dark points have predicted $T_{\rm eff}$ within 50 K of the spectroscopic value of 5\,730 K, and the others are outside this range.  The best-fit ($\chi^2 = 2.56$) isochrone of 12.5~Gyr is plotted 
 as the heavy curve.  The stellar mass is 0.88 \Msun, the radius is 1.23 \Rsun, and the \logg\ is 4.21, which is reasonably consistent with 
 the parameters derived from the stellar spectrum (Sect. \ref{spectrum}).
 
 The 68\% confidence intervals (based on $\Delta \chi^2$) for the posterior parameter values 
 are: \teff\ = 5\,750-5\,814 K, \logg\ = 4.20-4.25 dex,  \mstar = 0.87-0.91 \Msun, \rstar = 1.13-1.33 \Rsun, and an age of 11.5 - 13 Gyr (upper limit of isochrone models).  There is a slight tension between the spectroscopically derived parameters and other parameters, i.e. the errors do not overlap (Fig. \ref{F7}).  Using an [$\alpha$/Fe] = +0.2 grid the minimum $\chi^2$ increases the discrepancy and the model age increases beyond 13 Gyr. On the other hand, a slightly higher $T_{\rm eff}$ and $\log g$ would reconcile these estimates and yield a slightly younger age.  Regardless, these comparisons suggest a model-dependent age of at least 10 - 11~Gyr, i.e. at least as old as the Galactic disk itself \citep{Martig2016}.


  \targeta\ has a \vmag\ of 11.137\,$\pm$\,0.040 (Table~\ref{Tab:StarIdentifiers}). Applying the interstellar 
 extinction of 0.150\,$\pm$\,0.025 mag  found in Sect.~\ref{spectrum}, the de-reddened \vmag\ is 10.987\,$\pm$\,0.047 mag.  In order to calculate the stellar parameters including 
 its age, we apply the Bayesian PARAM 1.3  tool \citep{daSilva2006}\footnote{http://stev.oapd.inaf.it/cgi-bin/param\_1.3}. This tool accept as input the stellar \teff , the metallicity, the de-reddened visual magnitude, \vmag\, and the parallax. Using the de-reddened \vmag\ and the distance range determined in Sect.~\ref{star} (and converting those distances to parallaxes), we ran three separate models using our observed \teff\ and [Fe/H]  (Sect.~\ref{spectrum}). We find results between 8.8 Gyr and 11 Gyr, masses of 0.8 -- 0.89 \Msun\, radii between 0.85 \Rsun\ and 1.6 \Rsun\ and \logg\ between 4.46 and 3.96 (Table \ref{Table:Girardi}).  We then compare with the observed \logg\ ( Sect.~\ref{spectrum}), in order to assess which of the 3 distances better matches the spectroscopic parameters. Our data indicate \logg\ = $4.15 \pm 0.1$ dex. This would be indicative of a distance of 210 pc. The age would in this case be 10.770 Gyr and the mass of the star would be \mstar\ \about\ 0.9 \Msun\ but with a slightly larger \rstar\  of \about\ 1.3 \Rsun. We note here, however, that the error bars in this particular model are large. 
\begin{table*}
 \centering
 \caption{The output from the PARAM 1.3 models \citep{daSilva2006}. Input is \teff\ = 5730$\pm$~50 K, [Fe/H] = -0.53 $\pm$~0.05 dex and \vmag\ = 10.987\,$\pm$\,0.047.}
 \label{Table:Girardi}
\begin{tabular}{lrrrr}
\hline
\hline
Distance (pc)  &   Age (Gyr)  & \mstar (\Msun) & \rstar (\Rsun) & \logg (dex)  \\
\hline
\hline
190 $\pm$20 & $8.829 \pm 3.493$ & $0.809 \pm 0.022$ & $0.854 \pm 0.058$ &$4.456 \pm 0.057$ \\
210 $\pm$20 & $10.770 \pm 1.450$ & $0.861\pm 0.041$ & $1.275 \pm 0.356$ &$4.134 \pm 0.224$ \\
230 $\pm$20 & $11.035 \pm 0.609$ & $0.892 \pm 0.018$ & $1.591 \pm 0.081$ &$3.957 \pm 0.057$ \\
\hline
\end{tabular}
\end{table*}

If we use the stellar parameters derived from our model of the observed spectrum (\teff\, \logg\ and [Fe/H]) Sect. \ref{spectrum} as input to derive the mass and radius based only on the equations of \cite{Torres2010},  we find higher values of \mstar = 1.0\,$\pm$\,0.07\,\Msun, and \rstar = 1.4\,$\pm$\,0.14\,\Rsun. These 
equations of \citet{Torres2010} are based on the observed high precision \mstar\ and \rstar\ of 95 eclipsing binary stars of different luminosity classes where the masses and radii are known to better than 3\%, leading to a numerical relation based on the stellar parameters. It is, however, difficult to know how well these relations specifically describe \targeta. The number of stars in the generation of the numerical relation is small and of course not enough to generate "empirical" isochrones and the parameters derived in this way have to be treated with care. Specifically, the ages derived from the DSEP and PARAM 1.3 models indicate that a 1 \Msun\ star would already be evolving towards the white dwarf stage and the mass of \targeta\ must thus be lower. On the other hand our observation of a lower value for \logg\ than would be expected for a star with a \mstar\ < 1 \Msun\ is indicative of that the radius of \targeta\ should be larger than 1 \Rsun. 

Based on the above, we conclude, that all known facts are consistent with \targeta\ being an 0.86\,\Msun\ star that has begun to evolve off the main sequence, has a \rstar\ of 1.2-1.3 \Rsun, and thus with a very high age of the star. Our models are consistent with an age that is $\gtrsim$\,10 Gyrs, most likely being 10.8 Gyr or somewhat larger.

\section{Transit and RV joint modeling}
\label{Sect:Joint_Modeling}

We performed the joint fit of the photometric and RV data using the code \texttt{pyaneti}, a Python/Fortran software suite based on Marcov Chain Monte Carlo (MCMC) simulations (Barrag\'an et al., in preparation). The \ktwo\ photometry we analyzed  are subsets of the \targeta's light curve extracted by \citet{VandJohn2014}. Here we selected $\sim$ 7 hours of data points around each of the 13 transits observed by \ktwo\, and de-trended each transit using a second order polynomial fitted to the out-of-transit data points. The RV data set includes the 6 FIES and 12 HARPS-N measurements presented in Sect.~\ref{Sect:RV}.

We used the equations of \citet{Mandel2002} to fit the transit light curves and a Keplerian orbit to model the RV measurements. We adopted the Gaussian likelihood described by the equation
\begin{equation}
\begin{split}
\mathcal{L} = &
\left[
\prod_{i=1}^{n_{\mathrm{tot}}}
\left ( 2 \pi \sigma_i^2 
\right )^{- 1/2}
\right]
\exp
\left\lbrace
- \sum_{i=1}^{n_{\mathrm{tot}}} \frac{ \left( D_i - M_i \right)^2}{2 \sigma_i^2 }
\right\rbrace,
\label{eq:eq1}
\end{split}
\end{equation}
where $n_{\mathrm{tot}} = n_\mathrm{rv} + n_\mathrm{tr}$ is the number of RV and transit points, $\sigma_i$ is the error associated to each data point $D_i$, and $M_i$ is the model associated to a given $D_i$.
We fit the same parameters as in \citet{Barragan2016} to the light curve. For the orbital period ($P_\mathrm{orb}$), mid-time of first transit ($T_0$), impact parameter ($b$), planet-to-star radius ratio ($R_\mathrm{p}/R_\star$), RV semi-amplitude variation ($K$), and gamma velocity, we set uniform uninformative priors, i.e., we adopted rectangular distributions over given ranges of the parameters spaces.The ranges are $T_0=[7067.9708, 7067.9786]$ days for the mid-time of first transit, $P_\mathrm{orb}=[5.3503, 5.3514]$ days for the orbital period, $b=[0, 1]$ for the impact parameter, $R_\mathrm{p}/R_\star=[0, 1]$ for the planet-to-star radius ratio, $K=[0, 1000]$~\ms\ for the RV semi-amplitude variation, and $\gamma_\mathrm{FIES} = [-17,-15]$~\kms\ and $\gamma_\mathrm{HARPS-N} = [-17,-15]$~\kms\ for the systemic velocities as measured with FIES and HARPS-N, respectively. 

Given the limited number of available RV measurements and their error bars, we assumed a circular orbit ($e=0$). We adopted a quadratic limb darkening law and followed the parametrization described in \citet{Kipping2013}. To account for the \ktwo\ long integration time ($\sim$30 minutes), we integrated the transit models over 10 steps. The shallow transit and \ktwo's long cadence data do not enable a meaningful determination of the scaled semi-major axis ($a_\mathrm{p}/R_\star$) and limb darkening coefficients $u_1$ and $u_2$. We thus set Gaussian priors for the stellar mass and radius (Sect.~\ref{spectrum}) and constrain the scaled semi-major axis using Kepler's third law. We also used the online applet\footnote{Available at \url{http://astroutils.astronomy.ohio-state.edu/exofast/limbdark.shtml}.} written by \citet{Eastman2013} to interpolate \citet{Claret2011}'s limb darkening tables to the spectroscopic parameters of the host star (Sect.~\ref{spectrum}) and set Gaussian priors for the limb darkening coefficients $u_1$ and $u_2$ adopting 20\% conservative error bars. We explore the parameter space with 500 chains created randomly inside the prior ranges. The chain convergence was analyzed using the Gelman-Rubin statistics. The number of iterations required for the Marcov Chains to converge ("burn-in phase") uses $25,000$ more iterations with a thin factor of $50$. The posterior distribution of each parameter has $250,000$ independent data points.

We searched for evidence of an outer companion in the RV measurements by adding a linear trend $\dot{\gamma}$ to the Keplerian model fitted to the
RV data. The best fitting solution provide a linear trend of $\dot{\gamma}=-0.217\pm0.077~{\rm\,m\,s^{-1}\,d^{-1}}$  with a $\sim$3-$\sigma$ significance level. To assess if this
model is better, we have to compare it with the model without linear trend. When comparing models, the one with the largest
likelihood has to be preferred. At the same time, we have to check if we are not overfitting the number of parameters with the
Bayesian information criteria (BIC). This is defined as BIC = $ k \ln(n) - 2 \ln \mathcal{L}$, where $n$ is the number of data
points and $k$ the number of fitted parameters. The BIC penalizes the model with more fitted parameters. When comparing
models with different number of parameters, we have to prefer the one with the smallest BIC \citep{Gelman2003}. For our RV
measurements, the model with linear trend has $\ln \mathcal{L}_{\rm RV}$=78 and $\mathrm{BIC_{RV}}$=-144, while the model
without it gives $\ln \mathcal{L}_{\rm RV}$= 74 and $ \mathrm{BIC_{RV}} $= -139. We therefore conclude that the model with a linear
trend is favored.

\begin{figure}[!t]
 \centering
   \includegraphics[width=\linewidth]{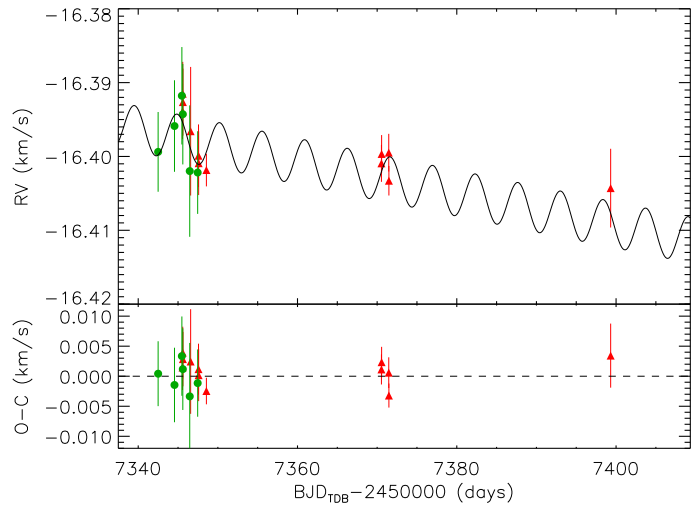}
   \caption{\emph{Upper panel}: FIES (green circles) and HARPS-N (red triangles) RV measurements versus time, following the correction for instrument offset. The best fitting Keplerian model with a linear trend is overplotted with a tick line. \emph{Lower panel}: Radial velocity residuals.}
   \label{F4}
 \end{figure}

\begin{figure}[!t]
 \centering
   \includegraphics[width=\linewidth]{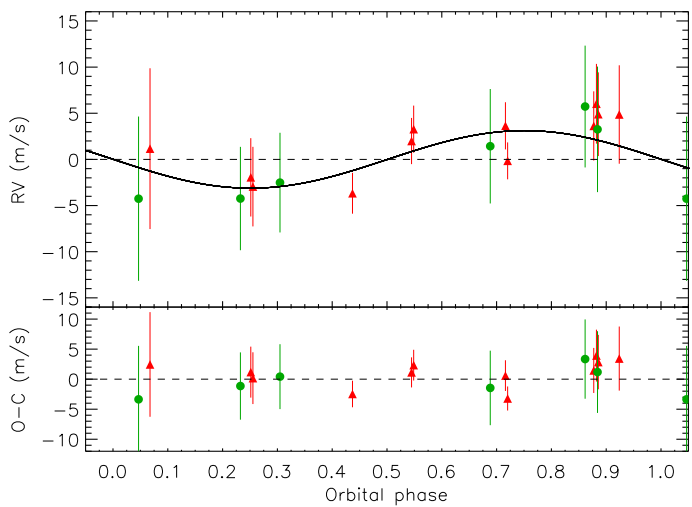}
   \caption{\emph{Upper panel}: FIES (green circles) and HARPS-N (red triangles) RV measurements phase-folded to the orbital period of \targetb, following the subtraction of the linear-trend. The best fitting circular model is overplotted with a tick black line. \emph{Lower panel}: Radial velocity residuals.}
   \label{F5}
 \end{figure}

The final parameters are given in Table~\ref{Table_Param}. They are defined as the median and 68\,\% credible interval of the posterior distribution for each parameter. We show the folded transit light curve in Fig.~\ref{F1} and the RV curves in Figs.~\ref{F4} and \ref{F5}.
\begin{figure}[!t]
 \centering
   \includegraphics[width=\linewidth]{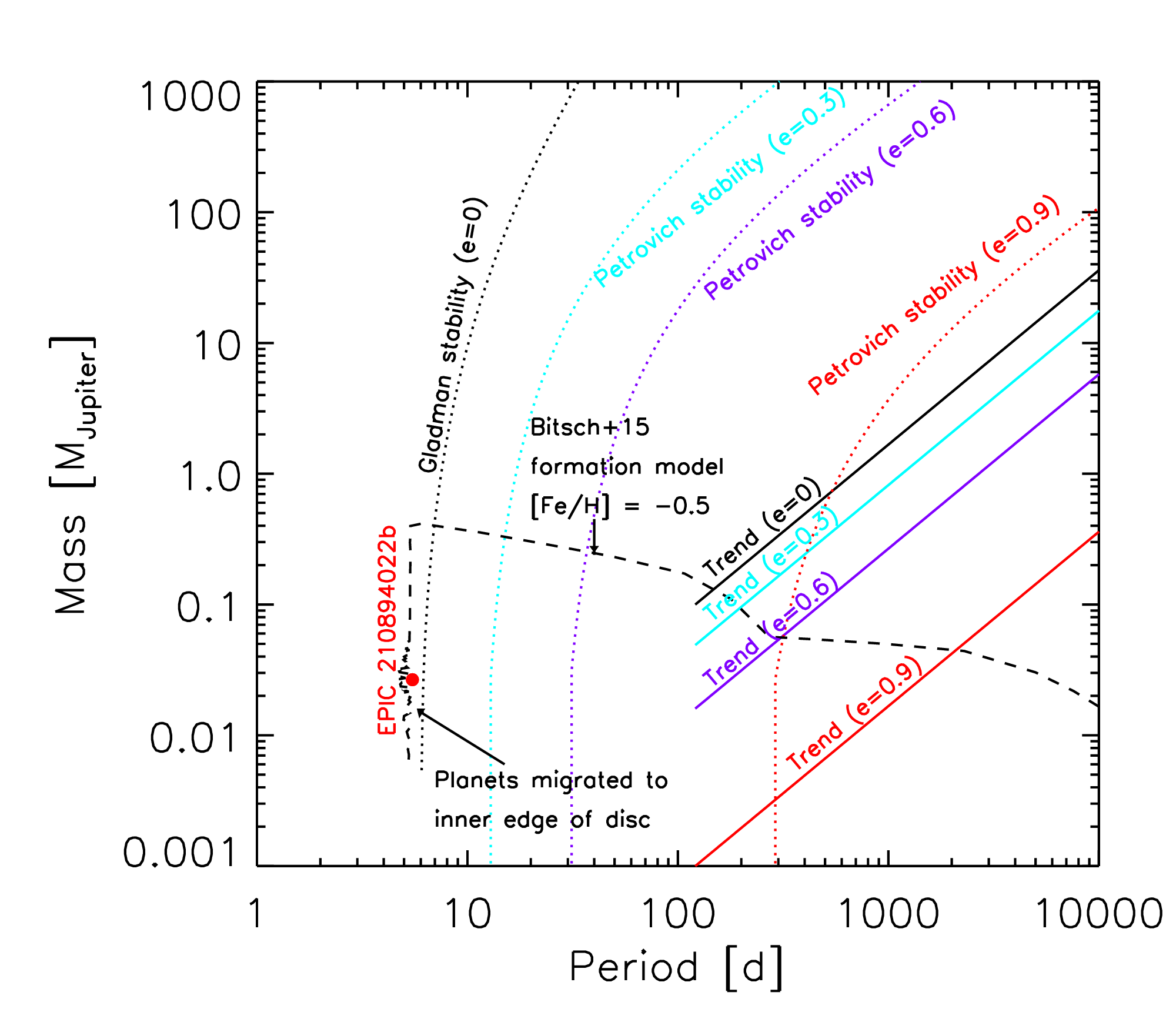}
   \caption{Constraints on the third body in the \targeta\ system. Solid lines show the minimum masses required to reproduce the RV trend, as a function of the third body's orbital period and eccentricity. Dotted lines show the maximum masses allowed for dynamical stability for these periods and eccentricities \citep{Gladman1993,Petrovich2015}. The dashed black line shows the final masses of planets produced in the planet formation model of \cite{Bitsch2015} with a metallicity $[Fe/H]=-0.5$. This model successfully forms super-Earths which migrate to \targetb's location at the inner disc edge, and predicts that the second body would have a mass of 20-50 \mearth. \targetb }
   \label{F6}
 \end{figure}

\section{Orbital Dynamics}
\label{Sect:Lund_Model}
The mass, orbital period and eccentricity of the body responsible for the RV trend can be constrained by requiring that the system is dynamically stable. Bodies too close, too massive, and on too eccentric orbits would result in an unstable system.

In Fig. \ref{F6}, we show for given periods of the outer body, the allowed mass ranges that are (a) large enough to generate the observed RV trend with P >120 d (above the solid lines); and (b) small enough to avoid dynamical instability (below the dotted lines). For an outer body on a circular orbit we use the criterion of \cite{Gladman1993} while for eccentric outer bodies we use \cite{Petrovich2015}.

We show results for four values of the outer body's eccentricity. If the outer body is on a circular orbit, it must be a gas giant planet or more massive, and the system is stable even for stellar-mass companions. If it is on a highly-eccentric orbit, gas giant planets at P\,\about\,1 yr are ruled out by dynamical stability. In this case, the outer planet may be a lower-mass planet on a close orbit (P\,\about\,1 yr) or a gas giant on a wider orbit (P >\,\about\,2 yr). Note that an eccentric orbit permits lower masses for the outer body, but this requires a specific alignment of the orbit with respect to the observer (edge-on orbit and pericentre pointing along the line of sight). In general, one can also place limits on what additional planets could be in a system between two known ones. For example, if the second planet is a Jupiter at 1 AU on a circular orbit, the separation is roughly 20 mutual Hill radii, meaning that one (or more) additional planets could be accommodated between the two planets.

We include a line in Fig. \ref{F6} that shows the final masses and orbital periods of planets formed in the planet formation model of \cite{Bitsch2015}. This model makes use of the accelerated core accretion rates by pebble accretion \citep{Lambrechts2014} and incorporates planet migration, meaning that the planets move through the disc as they form. Here, we use a simple power law disc model (with alpha viscosity parameter of 0.001) for the surface density and temperature following \cite{Ida2016} for sun-like stars to calculate the evolution of planets. We also make use of the metallicity measurements and evolve our planetary growth using a metallicity of [Fe/H] = -0.5.

The dashed line marks the final mass of planets as a function of their period as predicted by our simulations of planet formation. The vertical part of the line indicates that planets having a broad range of masses have migrated to the inner edge of the disc, where they stop their accretion. Our model here predicts that \targetb\ core has formed around 6 AU, i.e., beyond the water ice line.

The results from the simulations also indicate that the potential other companion in the system should be between 20\,-\,50 Earth masses, provided the planets evolved independently (they did not influence each other's growth and orbits). Follow-up observations of the planetary system can thus provide a deeper insight into the formation process of the planets in this system.
\section{Discussion and Summary}

The \targeta\ system is demonstrated to be a rare and important object among the plethora of transiting exoplanets that has been discovered by space missions in the last decade. Using adaptive optics imaging and statistical methods, and also detecting the RV signature of this planet we have confirmed the presence of a \rpb\ planet in a 5.35d orbit, as giving rise to the \ktwo\ transit signature. We find that the planet has a mass of \mpb. The periodic RV signal is overlaid on a trend that we identify with a second more massive object. The evidence for the planet \targetb\ are strong enough for us to say that it is confirmed, while we would require more data in order to confirm also the second body.

We believe this planet to be extremely old. The reasons for this is as follows: \emph{a}) The low but $\alpha$-rich metal content of \targeta. \emph{b}) This star has a very high space velocity of \about\ 145 \kms~making it a likely member of the thick disk population. \emph{c}) The modelling of the measured stellar parameters in Sect.~\ref{Age}. The best fit to the data is for a 0.86 \Msun\ star with a most likely age of 10.8 $\pm 1.5$ Gyr. The star appear to be beginning to move off the main-sequence as indicated by both the low value of \logg\ and the radius of the models that are most likely around $1.25 \pm 0.2 $ \Rsun. 

Different populations in the Galaxy can be traced through the abundance of the $\alpha$ elements, O, Mg, Si, S, Ca, and Ti. In this context we note that there 
are similarities between \targeta\ and the planet host star Kepler-444. The latter object is a metal poor low-mass solar-like star and one of the brightest stars to be observed with \kepler. By following this object during the 4 years of that mission, \cite{Campante2015} succeeded in detecting 5 transiting sub Earth-size planets in a compact system. They also could record the asteroseismic signature of the host star. Interpreting the seismic data allowed a high precision determination of mass (0.76\,\Msun), radii (0.75\,\Rsun) and age ($11.23 \pm$ 1 Gyr) for the host star by these authors. Kepler-444 has very similar space velocities (see Sect. \ref{star}) and $\alpha$ element abundance as \targeta\ does, something that indicate that both stars are bona-fide members of the thick disk population. It has also been suggested that Kepler-444 is a member of the Arcturus stream, a group of older iron poor stars that possibly originates from outside the Milky Way Galaxy. 

There exist data on a handful of other small size (super-Earth or Neptune class) planets, where there are also indications of high age. Kepler-10b and c \citep{Batalha2011,Fressin2011}, the first small planets confirmed by the \kepler\ mission, have been determined (asteroseismologically) to have an age of $11.9 \pm4.5$~Gyr. This system has been suggested to belong to the halo population \citep{Batalha2011}. The metallicity of the star is, however, higher than \targeta\ at [Fe/H] = $-0.15 \pm 0.03$. Also the error bars on the age are high, and no proper motions are available to kinematically determine the population of the star. The recently confirmed Kepler-510 system \citep{Morton2016} has a host star with a metallicity of [Fe/H] = $-0.35 \pm 0.1$ and an asteroseismic age of 11.8 Gyr \citep{Silvaaguirre2015}. While the planet (orbital period 19.6d) have a radius of \about~2.2 \rearth\ no mass of this object has as yet been determined. We point out in this context that future releases of the Gaia astrometric catalogue will alleviate this situation and allow for a kinematical determination of old host star populations. There is also the case of Kapteyn's star (GJ 191, LHS 29 or HD 33793), an M1 sub-dwarf star \citep{Gizis97} with a  [Fe/H] = $-0.86 \pm0.05$. It is kinematically classified as a halo star and is in fact the closest such object at a distance of only $3.91 \pm0.01$ pc. Two planets were detected in radial velocity measurements \citep{Anglada2014}, with 
periods of 48.6d and 121.5d and $m_p sin~i$ of 4.8 and 7.0 \mearth\ respectively. The age of the star is very likely older than 10 Gyr because of the low metallicity and the kinematics, but exactly how old it is can not be determined at this time. \cite{Robertson2015} used a somewhat different data set, almost as large as that of \cite{Anglada2014}, and concluded that the RV signature of Kapteyn-b very likely was caused by an activity signal coming from the star. \cite{Anglada2016}  analysed this latter data set and came to the conclusion that there is no activity signal but instead most likely the bona-fide planet-b is a real planet. This demonstrates the difficulty when one is working at the limit of the sensitivity of one's instrumentation.
While it is only the three objects Kepler-444, Kepler-510 and \targeta\ that have both confirmed planets and relatively well secured ages, very old stars appear to be as likely to possess planetary systems as younger systems, a not too surprising result. It is however more interesting in terms of what kind of planets form in early low-metallicity systems, as compared to the more recently formed systems where the metallicity would generally be higher.
 
It is clear that \targeta\ and its planet(s) is a welcome addition to Kepler-444 and Kepler-510. That \targeta\ is being abundant in $\alpha$ elements is interesting since the bulk of rocky planets consist of those elements \citep{Valencia2007,Valencia2010}.  Together with the 5 planets in the Kepler-444 system, Kepler-510 and possibly the other exoplanet systems described above, \targetb\ and its possible companion suggested here are among the oldest planets known to date. Assuming a radius of \rpb\, the planet has an average density of \rhob\ g cm$^{-3}$, placing it in the same class as far as geometrical size is concerned as \corseven\ and Kepler-10b. In this context it is indeed a super-earth and the planetary density appear similar to that of Venus and the Earth itself. The errors in $\rho_\mathrm{p}$ are, at the moment, however, large enough that it allow compositions that deviate from being truly "Earth-like" and more observations are required. It would have formed together with a star having a low metallicity, and more importantly at a very early epoch of our Galaxy. Although \targeta\ is also iron-poor, it is moderately $\alpha$-rich, in common with the planet host Kepler-444, which could be favourable for the formation of an Earth-like body. But we also have indications for a more massive planet in the same system. A number of studies so far have pointed out a correlation where metal-rich stars are more likely to harbour gas-giants (e.g. \cite{vf05,Mortier2013}), while the correlation appear to be missing for the sample of small planets discovered by \kepler\ \citep{Buchhave2012}. Having formed \about 5-6 Gyrs before the birth of the Solar System \targeta\ and its system carries information about the early stages of stellar and planetary formation in the Galaxy. It would therefore be very interesting to continue to study this system, primarily to confirm the presence of the second more massive planet and finding also its period. Finding more systems, similar to \targeta\ and Kepler-444, would allow us to begin to determine what the implications for planetary formation as a function of galactic age is.

\bibliographystyle{aa}
\bibliography{exomf12.bib}

\begin{acknowledgements}

We acknowledge the very constructive comments of an anonymous referee, which have improved our paper.

We thank the McDonald, NOT, TNG, Subaru staff members for their unique and superb support during the observations.  

Based on observations obtained \emph{a}) with the Nordic Optical Telescope (NOT), operated on the island of La Palma jointly by Denmark, Finland, Iceland, Norway, and Sweden, in the Spanish Observatorio del Roque de los Muchachos (ORM) of the Instituto de Astrof\'isica de Canarias (IAC); \emph{b}) with the Italian Telescopio Nazionale Galileo (TNG) also operated at the ORM (IAC) on the island of La Palma by the INAF - Fundaci\'on Galileo Galilei. 

This research made use of data acquired with the Carlos S\'anchez Telescope, operated at Teide Observatory on the island of Tenerife by the Instituto de Astrof\'\i sica de Canarias.

The research leading to these results has received funding from the European Union Seventh Framework Programme (FP7/2013-2016) under grant agreement No. 312430 (OPTICON).

M.F. and C.M.P. gratefully acknowledge the support of the Swedish National Space Board.

M.F. acknowledge the hospitality of the Instituto de Astrof\'\i sica de Canarias where the paper was written during a 2-month stay under a Jesus Serra Foundation fellowship.

D.G. gratefully acknowledges the financial support of the Programma Giovani Ricercatori -- Rita Levi Montalcini -- Rientro dei Cervelli (2012) awarded by the Italian Ministry of Education, Universities and Research (MIUR).

W.D.C., M.E. and P.M.Q. were supported by NASA grants NNX15AV58G, NNV16AE70G and NNX16AJ11G to The University of Texas at
Austin.

Sz.Cs. thanks  the  Hungarian  National  Research,  Development  and  Innovation  Office,  for  the  NKFIH-OTKA K113117 grant.

This work has made use of SME package, which benefits from the continuing development work by  J. Valenti and N. Piskunov and we gratefully acknowledge their continued support. This work has made use of the VALD database, operated at Uppsala University, the Institute of Astronomy RAS in Moscow, and the University of Vienna \citep{kupka2000,Ryabchikova2015}.

\end{acknowledgements}

\end{document}